\documentclass[]{piparticle-final}
\usepackage{graphicx}
\usepackage{amsmath}

\usepackage{cite} 
\usepackage{epstopdf} 

\begin{document}

\volume{7}               
\articlenumber{070007}   
\journalyear{2015}       
\editor{L. A. Pugnaloni}   
\reviewers{F. Vivanco, Dpto. de F\'isica, Universidad de Santiago de Chile, Chile.}  
\received{9 December 2014}     
\accepted{13 April 2015}   
\runningauthor{R. O. U\~nac {\it et al.}}  
\doi{070007}         
\title{Density distribution of particles upon jamming after an avalanche in a 2D silo}

\author{R. O. U\~nac,\cite{inst1}\thanks{E-mail: runiac@unsl.edu.ar}\hspace{0.5em}  
        J. L. Sales,\cite{inst2}\hspace{0.5em}  
        M. V. Gargiulo,\cite{inst2}\hspace{0.5em}  
        A. M. Vidales\cite{inst1}\thanks{E-mail: avidales@unsl.edu.ar}}

\pipabstract{
We present a complete analysis of the density distribution of particles in a two dimensional silo after discharge. Simulations through a pseudo-dynamic algorithm are performed for filling and subsequent discharge of a plane silo. Particles are monosized hard disks deposited in the container and subjected to a tapping process for compaction. Then, a hole of a given size is open at the bottom of the silo and the discharge is triggered. After a clogging at the opening is produced, and equilibrium is restored, the final distribution of the remaining particles at the silo is analyzed by dividing the space into cells with different geometrical arrangements to visualize the way in which the density depression near the opening is propagated throughout the system. The different behavior as a function of the compaction degree is discussed.
}

\maketitle

\blfootnote{
\begin{theaffiliation}{99}
   \institution{inst1} Departamento de F\'{\i}sica, Instituto de F\'{\i}sica Aplicada (UNSL-CONICET), Universidad Nacional de San Luis, Ej\'ercito de los Andes 950, D5700HHW San Luis, Argentina.
   \institution{inst2} Departamento de Geof\'{\i}sica y Astronom\'{\i}a. Facultad de Ciencias Exactas F\'{\i}sicas y Naturales. Universidad Nacional de San Juan, Mitre 396 (E), J5402CWH San Juan, Argentina.
\end{theaffiliation}
}

\section{Introduction}
Numerous studies have investigated the flow of granular materials (such as glass beads, aggregates or minerals, among others) through hoppers of various geometries \cite{ref1,ref2,ref3,ref4,ref5,ref6,ref7,ref8}. Commonly, the silo is initially filled with the granular material at a given height. Then, the outlet of the silo is opened and the mass flow rate is recorded as a function of time. These experiments provide useful information on the relation between the flow rate and different geometrical and physical properties of the system, such as the size and shape of the particles and the outlet, the presence of friction forces and density fluctuations \cite{ref9}. Empirically, the flow rate is determined by the known Beverloo's equation which depends, among other variables, on the apparent density in the immediate neighborhood of the outlet region of the silo. Furthermore, in the derivation of the Beverloo's equation, it is assumed that the region primarily affected 
by the discharge is near the outlet and 
has an effective diameter of the order of the width of it, sometimes referred to as Beverloo's diameter \cite{ref2}.

Simulations have provided details about the granular flow that are not accessible to experiments such as the influence of frictional parameters between particles, particle shape, stress chains and the statistics of the arches formed during a jamming process \cite{ref3,ref4,ref10,ref11,ref12}.

When the size of the flowing particles is in the order of the width of the aperture, jamming can take place, and the particles stop flowing unless additional energy is provided to the system. Many practical problems are caused by jamming at the outlet of hoppers used in production lines, where it is necessary to maintain a constant flow of material. Analogous problems occur in the storage of raw materials in silos, especially after a certain period of accumulation or under manipulation operations that may cause a change in the packing fraction. Thus, when it is necessary to empty the silo, the material does not flow, either due to the presence of unwanted moisture or because of the compaction of grains sealing the outlet \cite{ref13,ref14,ref15,ref16,ref17}. These problems are caused by particle arches that form at the outlet. Authors in Ref. \cite{ref9} report data suggesting temporal oscillations in the packing fraction near the outlet. Those oscillations had a frequency around 2 Hz. This result is 
important because it is directly related to the likelihood of jamming in the silo \cite{ref4,ref5}. Other authors performed flow experiments using metal disks in a two dimensional hopper in order to study the statistical properties of the arches forming at the outlet \cite{ref5}. Authors in Ref. \cite{ref7} found experimentally a linear variation for the number of particles forming an arch with the outlet size in a 2-D silo.

There are many works concerning the study of packing density in flowing granular materials out of a hopper. In particular, those dealing with the presence of density waves and density fluctuations in the bulk of the silo \cite{ref18,ref19}. Others focus on the density distribution during discharge and analyze the change of the density between the stagnant zone and the flowing zone \cite{ref20}.

In a recent work \cite{ref12}, authors have studied the jamming occurring in the flow through small apertures for a column of granular disks via a pseudo-dynamic model. The effect that the preparation of the granular assembly has on the size of the avalanches was investigated. To this end, packing ensembles with different mean packing fractions were created by tapping the system at different intensities. This work succeeded in demonstrating that, for a given outlet size, different mean avalanche sizes are obtained for deposits with the same mean packing fraction that were prepared with very different tap intensities. Nevertheless, a complete characterization of the density of particles inside the column, both before and after the discharging process, was left out.

For all above, a study of the variation of density near the outlet of a silo, before and after the discharge in the presence of a jamming, is important to relate it to the subsequent behavior of the system during the restart of the flow.

\section{Simulation procedure}

The implementation and description of a simulation algorithm using a pseudo-dynamic code has been developed in numerous studies since the liminal work of Manna and Khakhar \cite{ref12,ref21,ref22,ref23}. Assuming inelastic massless hard disks that will be deposited in a 2D die simulating a silo, the pseudo-dynamics will consist in small falls and rolls of the grains until they come to rest by contacting other particles or the system boundaries. We use a container formed by a flat base and two flat vertical walls. No periodic boundary conditions are applied.

The deposition algorithm consists in choosing a disk in the system and allowing a free fall of length $\delta$ if the disk has no supporting contacts, or a roll of arc-length $\delta$ 
over its supporting disk if the disk has one single supporting contact \cite{ref12,ref21,ref22,ref24}. Disks with two supporting contacts are considered stable and left in their positions. If during a fall of length $\delta$ a disk collides with another disk (or the base), the falling disk is put just in contact and this contact is defined as its \textit{first supporting contact}.
Analogously, if in the course of a roll of length $\delta$, a disk collides with another disk (or a wall), the rolling disk is put just in contact. If the \textit{first supporting contact} and the second contact are such that the disk is in a stable position, the second contact is defined as the \textit{second supporting contact}; otherwise, the lowest of the two contacting particles is taken as the \textit{first supporting contact} of the rolling disk and the \textit{second supporting contact} is left undefined. If, during a roll, a particle reaches a lower position than the supporting particle over which it is rolling, its \textit{first supporting contact} is left undefined (in this way, the particle will fall vertically in the next step instead of rolling underneath the first contact). A moving disk can change the stability state of other disks supported by it; therefore, this information is updated after each move. The deposition is over once each particle in the system has 
both supporting contacts defined 
or is in contact with the base (particles at the base are supported by a single contact). Then, the coordinates of the centers of the disks and the corresponding labels of the two supporting particles, wall or base are saved for analysis.

An important point in these simulations is the effect that the parameter $\delta$ has in the results since particles do not move simultaneously but one at a time. One might expect that in the limit $\delta \rightarrow 0$, we should recover a fairly "realistic" dynamics for a fully inelastic non-slipping disk dragged downwards at constant velocity. This should represent particles deposited in a viscous medium or carried by a conveyor belt. We chose $\delta=0.0062 d$ (with $d$ the particle diameter) since we have observed that for smaller values of $\delta$, results are indistinguishable from those obtained here \cite{ref12,ref24}.

It is worth saying that the pseudo-dynamic algorithm used here allows the final configurations obtained after each tapping to be completely static, given that each disk is supported by other two disks as required by the equilibrium conditions in the model. In this way, one can follow the history of formation of the packing and, thus, the occurrence of the arches. This leads to a straightforward definition of the arches in the system, as we will see below. On the other hand, this algorithm is faster in the generation of a given ensemble of particles and, the subsequent tapping process, than the DEM one. The ones above are the most important advantages of the algorithm that justify our choice.

Because arch formation has been identified as a potential cause for segregation in non-convecting systems \cite{ref25,ref26,ref27}, we have centered previous research on detecting arches and analyzing their behavior and distribution in piles and jammed silos \cite{ref23,ref28}. Indeed, identification of arches is a rather complex task and the results presented in those works were novel and original at that time. We recommend to those readers interested in the characterization of the arches formed before and after the discharge of a 2D silo filled with disks to address our previous work in Ref. \cite{ref12}.

Arches are sets of mutually stabilizing particles in a static granular sample. In our pseudo-dynamic simulations we first identify all mutually stable particles and then find the arches as chains of particles connected through these mutual stability contacts. Two disks A and B are said mutually stable if A is the left supporting particle of B and B is the right supporting particle of A, or vice versa. Since the pseudo-dynamics rest on defining which disk is a support for another disk during the deposition, this information is available in our simulations. Chains of mutually stable particles can, thus, be found straightforwardly. These chains can have, in principle, any size starting from two particles. Details on the properties of arches found in pseudodynamic simulations can be found in previous works \cite{ref7,ref24,ref28}.

\section{Filling and emptying the silo}

As explained above, the aim of the present work is to analyze the density patterns of particles inside a silo after its discharge and as a function of the compaction degree before the avalanche event.

We first need to prepare packings at reproducible initial packing fractions. To achieve this, we have chosen a well known technique to generate reproducible ensembles of packings \cite{ref12}. Thus, we use a simulated tapping protocol (see below) to generate sets of initial configurations that have well defined mean packing fractions. The simulations are carried out in a rectangular box of width $24.78d$ containing $1500$ equal-sized disks of diameter $d$. Initially, disks are placed at random in the simulation box (with no overlaps) and deposited using the pseudo-dynamic algorithm. Once all the grains come to rest, the system is expanded in the vertical direction and randomly shaken to simulate a vertical tap. Then, a new deposition cycle begins. After many taps of given amplitude, the system achieves a steady state where all characterizing parameters fluctuate around equilibrium values independently of the previous history of the granular bed. The existence of such "equilibrium" states has been previously 
reported in experiments \cite{ref29}.

The tapping of the system is simulated by multiplying the vertical coordinate of each particle by a factor $A$  (with $A>1$). Then, the particles are subjected to several (about 20) Monte Carlo loops where positions are changed by displacing particles a random length $\Delta r$ uniformly distributed in the range $0<\Delta r<A-1$. New configurations that correspond to overlaps are rejected. This disordering phase is crucial to avoid particles falling back again into the same positions. Moreover, the upper limit for $\Delta r$ (i.e., $A-1$) is deliberately chosen such that a larger tap promotes larger random changes in the particle positions. For each value of $A$ studied, $10^{3}$ taps are carried out for equilibration followed by $5\times10^{3}$ taps for production. $500$ deposited configurations are stored which are obtained by saving every ten taps during the production run after equilibration. These deposits will be used later as initial conditions for the discharge and flow through an opening \cite{
ref12}.

It is worth mentioning that friction is not considered in our present model, neither between particles nor between particles and the walls (convection is not present \cite{ref12}). The key feature to stabilize the particles inside the silo is the sticking to the base of the bottom particles. This ensures the achieving of a final static system. As one would expect, the presence of friction would affect the packing density. In this sense, we have already studied the effect of a kind of adhesion force like capillary forces in the packing fraction of tapped columns of disks in a previous work \cite{ref28}.

For each deposit generated, we trigger a discharge by opening an aperture of width $D$ relative to the diameter $d$ of the disk in the center of the containing box base. Grains will flow out of the box following the pseudo-dynamics until a blocking arch forms or until the entire system is discharged (with the exception of two piles resting on each side of the aperture). During the dynamics, disks that reach the bottom and whose centers lie on the interval that defines the opening will fall vertically (even if the surface of the disk touches the edge of the aperture). This prevents the formation of arches with end disks sustained by the vertical edge of the orifice. After each discharge, we record the final arrangement of the grains left in the box.

\begin{figure}
\begin{center}
\includegraphics[width=0.9\columnwidth,trim=0cm 0.3cm 0cm 0.5cm,clip]{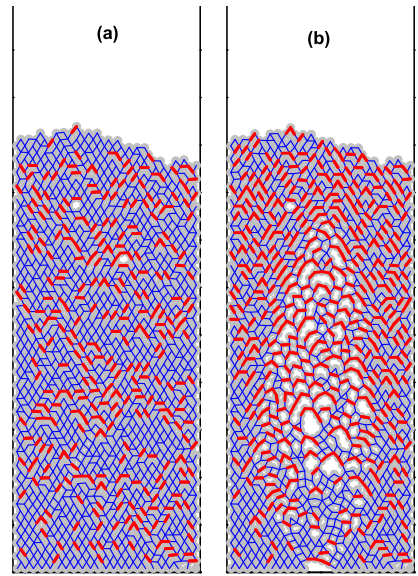}
\end{center}
\caption{Snapshots of the packed particles inside the plane silo, (a) before the discharge and (b) after it. The case corresponds to $A=1.1$ and $D=2.5$. Thin blue lines indicate contacts between particles and thick red lines indicate the arches.} \label{figure1}
\end{figure}

One single discharge attempt is carried out for each initial deposit. This allows us to assure that the initial preparation of the pack belongs to the ensemble of deposits corresponding to the steady state of the particular tap intensity chosen. To illustrate the initial and final packing configurations inside the silo, we present two snapshots in Fig. \ref{figure1} showing the particles before, part (a), and after, part (b), the discharge for the case $A=1.1$ and $D=2.5$.

In a previous work, we have focused on the effect that the preparation of the granular assembly has on the size of the avalanches obtained. In the next section, we present the analysis of the corresponding packing densities before and after the discharge.

\section{Density sampling}

\begin{figure}
\begin{center}
\includegraphics[width=0.9\columnwidth,trim=0cm 0cm 0cm 0cm,clip]{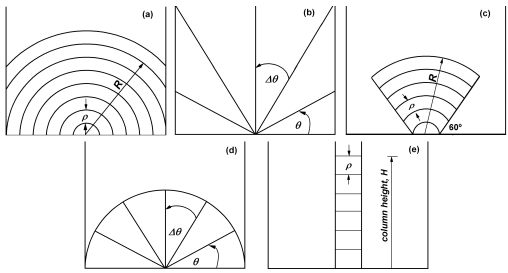}
\end{center}
\caption{Sketch of the geometry of the cells used to measure the density of particles inside the silo. (a) Circular ring sectors of thickness $\rho$,  centered at the middle point of the base. (b) Radial angular sectors starting at the center of the base, spanning all the packing. (c) Sectors as those in part (a) but delimited by two symmetrical lines at $60^{\circ}$ respect to the horizontal and a maximum radius of half width of the base. (d) Sectors as in part (b) but limited by a semicircle with radius of half width of the base. (e) Column of rectangular sampling sectors, with height $\rho$, and width equal to the outlet size.} \label{figure2}
\end{figure}

To measure the density patterns, we perform the analysis by dividing the packing space into cells, following different geometrical criteria, and for $A=1.1$, $1.5$, $2.0$, and $D=2.5$. The first measurements are performed over all the particles inside the silo. We choose two particular arrangements for the sampling cells, i.e., circular ring sectors centered at the base, the maximum radius for the ring being five times the width of the base, and radial angular sectors starting at the center of the base, where the opening of the sector is $\Delta \theta$, with the inclination $\theta$ measured from the horizontal axis. Figure \ref{figure2} (a) and (b) illustrate the cases. On the other hand, and to especially focus on the region near the outlet, we define in addition three different cell configurations. They are shown in Fig. \ref{figure2} (c)-(e). The details are depicted in the figure caption.

In all the sampling cell analysis, the criterion for density calculation was the same. The particles whose centers fall into a given sector are counted, and that number is then divided by the sector area, giving thus the particle density in the sector. We measure the density corresponding to the initial packing and for the packing array after the discharge for each one of the independent configurations and obtain the mean values of the density over those configurations presenting clogging. This was repeated for for all the geometries used.

\section{Results and discussion}

\begin{figure}
\begin{center}
\includegraphics[width=0.98\columnwidth,trim=0.3cm 0.7cm 0.7cm 0.6cm,clip]{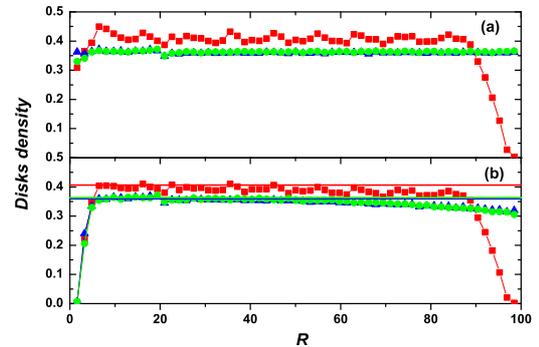}
\end{center}
\caption{Particle density vs. the distance of the ring to the center, using the cell depicted in Fig. \ref{figure2}(a). Here $\rho=d$. (a) Initial packing behavior. (b) Final stage after discharge. $A=1.1$ (red squares), $1.5$ (blue up triangles) and $2.0$ (green circles).} \label{figure3}
\end{figure}

In Fig. \ref{figure3} we present the results for the particle density as a function of the distance of the ring to the center (see Fig. \ref{figure2} (a)). The thickness of each ring is equal to $1.61$ a.u., i.e., a particle diameter. In part (a) of the figure we plot the behavior of the density for the initial packings, before the discharge of the silo is performed. As expected, a constant behavior is observed as the radius increases. As also reported elsewhere \cite{ref28}, the average packing density is larger for $A=1.1$ than for $A=1.5$ or $2.0$. The oscilations observed for $A=1.1$ are due to the presence of order in the packing structure \cite{ref24}. This order is virtually absent for higher tapping intensities. The decrease of the density at small distances from the center, especially for $A=1.1$ and $2.0$, is a purely geometric effect related to the small area covered by the smaller rings and the particular disposition of the disks at the base. On the other hand, the decrease observed at 
large 
distances for $A=1.1$ is because the initial arrangement of particles presents a free surface which is tilted.

After the discharge (Fig. \ref{figure3} (b)), the disk density steeply drops inside the region close to the outlet, putting in evidence the size of the concave hole formed after the avalanche. Besides, the density slightly falls with height, for all amplitudes. To highlight this effect, we draw three lines corresponding to the mean densities at the initial state (Fig. \ref{figure3} (a)). This effect is due to the lower density of packing near the walls and it will be explained later. Here again, the oscillations for $A=1.1$ are associated to the order in the packing structure. A similar analysis can be done taking a different thickness for the rings, giving results qualitatively equal to the ones shown in Fig. \ref{figure3}.

\begin{figure}
\begin{center}
\includegraphics[width=0.98\columnwidth,trim=0.8cm 0.6cm 0.7cm 0.6cm,clip]{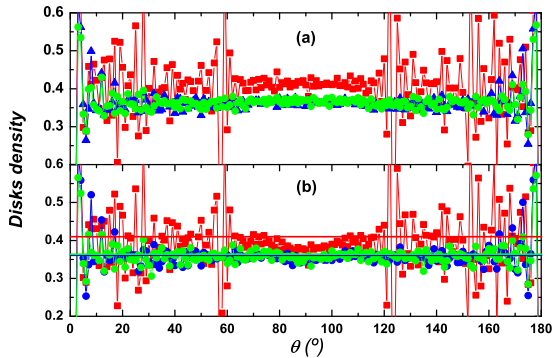}
\end{center}
\caption{Particle density vs. the inclination angle of the sector respect to the horizontal, for three values of $A$. (a) Initial packing behavior. (b) Final stage after discharge. $A=1.1$ (red squares), $1.5$ (blue up triangles) and $2.0$ (green circles). The three horizontal lines indicate the initial average densities for the three amplitudes.} \label{figure4}
\end{figure}

In Fig. \ref{figure4} we present the results for particle density as a function of the inclination angle of the sector respect to the horizontal for three values of $A$. Part (a) corresponds to the initial packing structure and part (b) shows the state after discharge. The opening angle in each sector is $1^{\circ}$. The horizontal axis represents the angle of the most inclined side of the sector, as indicated in Fig. \ref{figure2} (b). The peaks for $A=1.1$ in both plots are associated with the ordered structure, showing important increments of the density for $60^{\circ}$ and $120^{\circ}$ and, less important, for $30^{\circ}$ and $150^{\circ}$. When increasing the angular sector width, $\Delta \theta$, peak amplitude decreases, virtually disappearing, keeping the qualitative behavior shown in Fig. \ref{figure4}. The average density for $A=1.1$ is slightly higher, as expected.

After discharge, a dip at the central part of the curves appears as an evident consequence of the created hole, and showing the presence of a central region around the hole in which the density is reduced. This reduction is more pronounced for $A=1.1$ because the mean avalanche size for that tapping intensity is greater \cite{ref12} and the affected region is approximately between $60^{\circ}$ and $120^{\circ}$ for all amplitudes. At the bottom part of the figure, we plotted three lines indicating the initial average densities for the three amplitudes to better visualize the density change.

Although useful to have an overview for density behavior, information is lost when averaging over sectors spanning all packing configuration. For that reason, and to analyze in more detail the region close to the outlet of the bidimensional silo, we implement three new arrangements for the sectors to perform the density analysis, as indicated in Fig. \ref{figure2} (c)-(e). Fig. \ref{figure2} (c) shows a scheme for sectors similar to those in part (a) but delimited by two symmetrical lines at $60^{\circ}$ respect to the horizontal and a maximum radius of $20$ a.u. (half-width of the container). Part (d) sketches the same sectors as in part (b) but limited by a semicircle with radius $20$ a.u.. Finally, part (e) in the same figure, shows a column of rectangular sampling sectors, each one with height $\rho$, and width equal to the outlet size.

\begin{figure}
\begin{center}
\includegraphics[width=0.98\columnwidth,trim=0cm 0.5cm 0cm 0.5cm,clip]{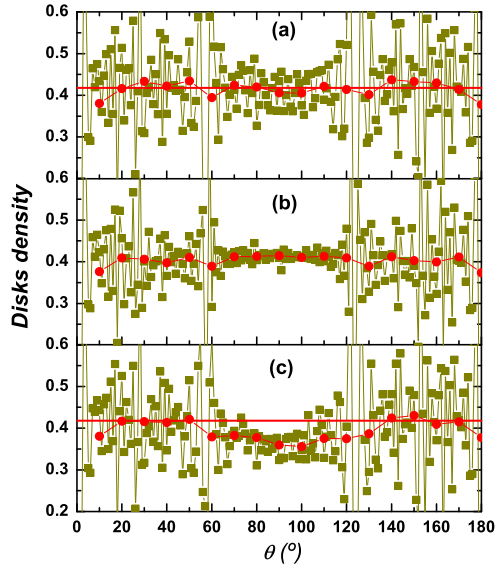}
\end{center}
\caption{Density average vs. inclination angle for $A=1.1$. Squares correspond to $\Delta \theta=1^{\circ}$ and circles for $\Delta \theta=10^{\circ}$. Upper: initial packing structure. Middle: initial density but for the sectors without the boundary semicircle, for comparison. Bottom: final state after the discharge for the system in the upper part. The horizontal line indicates the initial average density for $\Delta \theta=10^{\circ}$.} \label{figure5}
\end{figure}

In Fig. \ref{figure5} we present the results for $A=1.1$ for the density averaged over sectors as in Fig. \ref{figure2} (d). The upper part shows the density for the initial packing structure as a function of the inclination angle of the sector with $\Delta \theta=1^{\circ}$ (squares) and $\Delta \theta=10^{\circ}$ (circles). For comparison, the middle part of the figure also shows the initial density but for the sectors without the boundary semicircle. By circumscribing the density calculation to a semicircle centered at the outlet and whose radius is the half-width of the system, the ordered structure of the bottom part of the initial packing is more evident (peak occurrence) inside sectors from $60^{\circ}$ to $120^{\circ}$.

Analyzing the final state (bottom part of Fig. \ref{figure5}), a visible decrease in density around the outlet and also a large disorder is observed, especially between $60^{\circ}$ and $120^{\circ}$. Outside this range, the structure of the packing seems virtually unchanged, thus revealing the non-avalanche area. As before, the lines indicate the initial average density for the packing.

\begin{figure}
\begin{center}
\includegraphics[width=0.98\columnwidth,trim=0cm 0.5cm 0cm 0.5cm,clip]{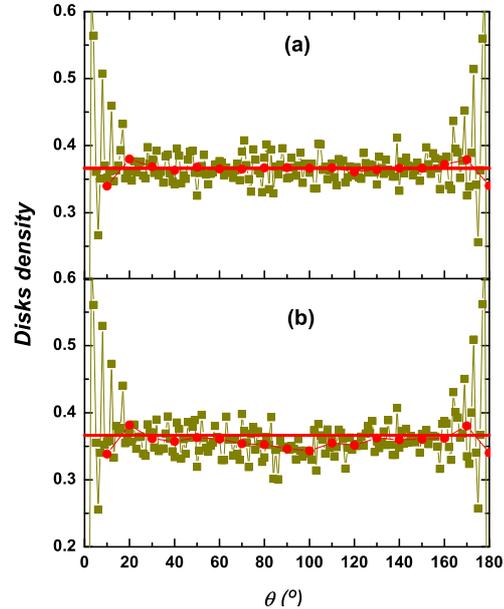}
\end{center}
\caption{Density vs. inclination angle as in Fig. \ref{figure5} (a) and (c), but for $A=1.5$, $\Delta \theta=1^{\circ}$ (squares) and $\Delta \theta=10^{\circ}$. The horizontal line indicates the initial average density for $\Delta \theta=10^{\circ}$.} \label{figure6}
\end{figure}

In Fig. \ref{figure6} we show the results for $A=1.5$ with $\Delta \theta=1^{\circ}$ (squares) and $\Delta \theta=10^{\circ}$ (circles) . It is clear from part (a) of the figure that density oscillations are much smaller than those for $A=1.1$, even for $\Delta \theta=1^{\circ}$. This proves the lack of order in the packing structure, except for those sectors close to the base of the packing.

After discharge, the density does not change substantially throughout the bulk, except for a slight decrease in the vicinity of the outlet orifice. This is shown in Fig. \ref{figure6} (b) and it is more evident for $\Delta \theta=10^{\circ}$. A slight density modulation can also be observed.

As a partial conclusion, we can say that the density of the packing after discharge retains the look of the original grain disposition, which is consistent with the disorder obtained for that tapping amplitude, i.e., lowering of the packing fraction \cite{ref12}. The results and conclusions for $A=2.0$ are quite similar to those for $A=1.5$.

\begin{figure}
\begin{center}
\includegraphics[width=0.98\columnwidth,trim=0cm 0.5cm 0cm 0.5cm,clip]{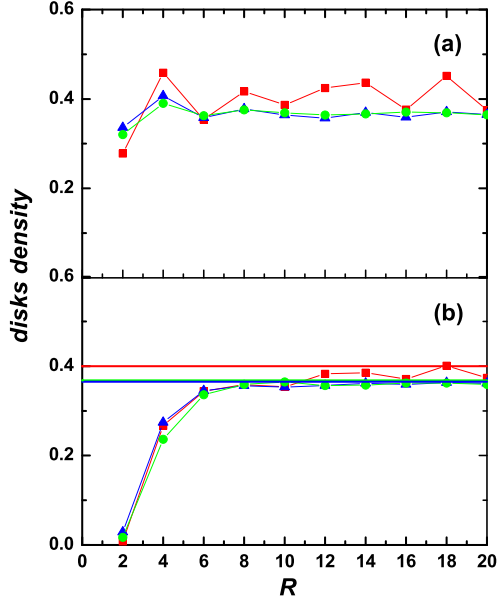}
\end{center}
\caption{Results for the density of particles averaged over sectors as in Fig. \ref{figure2} (c), for $A=1.1$, $1.5$ and $2.0$, and $\rho=2$ a.u.. (a) Initial packing. (b) After the discharge. $A=1.1$ (red squares), $1.5$ (blue up triangles) and $2.0$ (green circles). The three horizontal lines indicate the initial average densities for the three amplitudes.} \label{figure7}
\end{figure}

In Fig. \ref{figure7} the results for density are plotted averaging over sectors as in Fig. \ref{figure2} (c), for $A=1.1$, 1.5 and $2.0$, and $\rho=2$ a.u.. As in previous cases, the initial packing shows the periodicity related to the ordered structure for small $\rho$ (not shown here) and small $A$. As $\rho$ increases, fluctuations become smaller and a practically constant value for packing density can be observed as a function of the distance to the bottom center of the silo.

After discharge (Fig. \ref{figure7} (b)), a sudden decrease in density is observed up to a height equivalent to the estimated Beverloo's diameter, i.e., in our present case, $4$ a.u. Two regions can be distinguished. One corresponding to a radius of $2$ a.u., where the density is zero, and the other, where the density increases rapidly, almost reaching the initial bulk value.

For $A=1.5$, the fluctuations are only present for small $R$, near the base. After discharge, the behavior is similar to the case $A=1.1$, but with the presence of much less fluctuations.

For all amplitudes, the depression in density is confined inside the cone subtended by the lines at $60^{\circ}$. Compare Fig. \ref{figure7} with Fig. \ref{figure4} to see that the sectors corresponding to the stagnant zone keep the initial disk density. A similar result is obtained for $A=2.0$.

\begin{figure}
\begin{center}
\includegraphics[width=0.98\columnwidth,trim=0cm 0.5cm 0cm 0.5cm,clip]{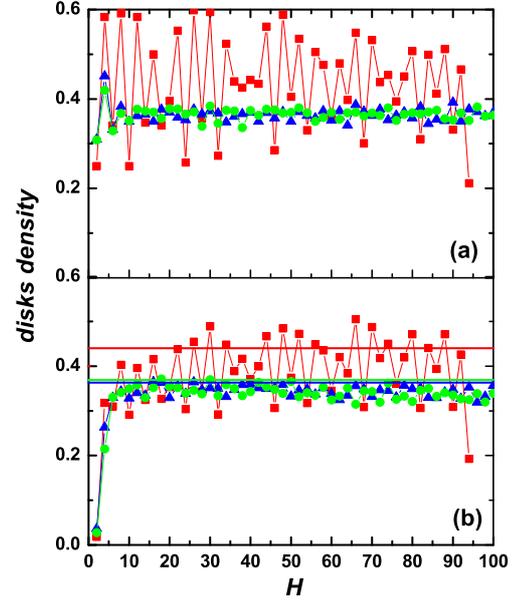}
\end{center}
\caption{Density results on the column vs. height $H$ belonging to sectors with thickness $\rho=2$ a.u. ($1.24d$). $A=1.1$ (red squares), $1.5$ (blue up triangles) and $2.0$ (green circles). The three horizontal lines indicate the initial average densities for the three amplitudes.} \label{figure8}
\end{figure}

Finally, we performed the density analysis on the column made by sectors with a given thickness $\rho$, as indicated in Fig. \ref{figure2} (e). Figure \ref{figure8} shows the results for $A=1.1$, $1.5$ and $2.0$, and $\rho=2$ a.u.. In part (a) of the figure, which corresponds to the initial packing structure, the density of disks remains constant with height for different $A$. Fluctuations are again related to the ordered structure and decreases for greater $A$. In the case $A=1.5$, as known, the initial configuration is more disordered than for $A=1.1$ (less fluctuations even for small $\rho$). The mean density of the packing coincides with the corresponding values in previous figures.

\begin{figure}
\begin{center}
\includegraphics[width=0.9\columnwidth,trim=0cm 0cm 0cm 0cm,clip]{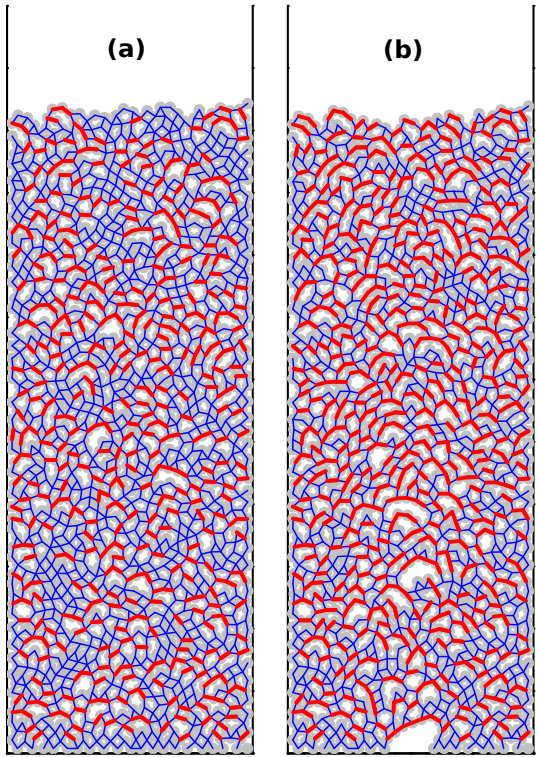}
\end{center}
\caption{Snapshots of the particles inside the silo, (a) before the discharge and (b) after it. The case corresponds to $A=1.5$ and $D=2.5$. Thin blue lines indicate contacts between particles and thick red lines indicate the arches.} \label{figure9}
\end{figure}

After the avalanche (Fig. \ref{figure8} (b)), a change in density is observed up to a height of about $25$ to $30$ a.u.. Up to the order of $40$ a.u., fluctuations decrease significantly, while for greater heights (far from the outlet) only a slight decrease of those fluctuations is observed. This is probably related to the formation of arches in the structure that prevents the occurrence of internal avalanches at greater heights. Figure \ref{figure1} and Fig. \ref{figure9} illustrate this point. Those figures show the snapshots before and after the discharge for $A=1.1$ and $A=1.5$, respectively. There, the arches formed by the particles are indicated with thick segments. In parts (b), a loose packing with the presence of arches is the resulting structure after the avalanche \cite{ref12}.

The density drop in Fig. \ref{figure8} (b) coincides with that in Fig. \ref{figure7} (b) and its extent is related to the region mainly affected by the discharge, which is of the order of one Beverloo's diameter ($4$ a.u.). Besides, a decrease of about $5\%$ to $10\%$ in the average bulk density is observed, in coincidence with Fig. \ref{figure3} (b).

\begin{figure}
\begin{center}
\includegraphics[width=0.98\columnwidth,trim=0cm 0.5cm 0cm 0.5cm,clip]{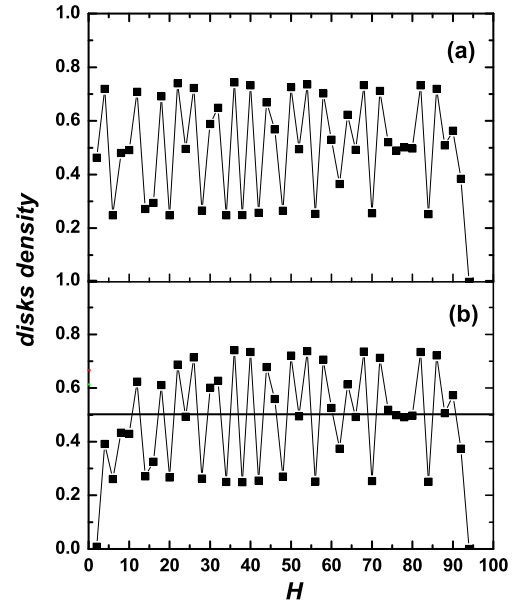}
\end{center}
\caption{Density results as in Fig. \ref{figure8} but $A=1.05$. The horizontal line indicates the initial average density.} \label{figure10}
\end{figure}

The analysis of the column for $A=1.05$ presents some interesting features comparing to the preceding results. Figure \ref{figure10} shows the results for that amplitude. First, we observe that the density fluctuations vs. height in the arrangements (both initial and final) are higher, reinforcing the fact that fluctuations increase with decreasing $A$. This is related with the higher order structure.

On the other hand, up to a height of $25$ to $30$ a.u. (which involves not only the empty vault area but the rarefaction caused by the avalanche), the density does not fluctuate for $A=1.5$ and $2.0$, while it does fluctuate for $A=1.05$ and $1.1$. This is because these latter arrangements are more compact and ordered, allowing propagation of the decreased density with a certain periodicity (see Fig. \ref{figure1}). As before, the radius of the empty vault coincides with the Beverloo assumption.

It is noteworthy that the decrease in the average density in the bulk after the discharge is only noticed when analyzing the data in a column or in the circular sectors of Fig. \ref{figure2} (a), not in the other cases. This would indicate that the main contribution to the avalanche is done by the particles just above the hole.

\section{Conclusions}

According to the results shown so far, it can be said that there is a clear area of rarefaction in the packing column after the avalanche discharge of the silo. This area is focused on the outlet opening. This lower density region has been analyzed with different geometries that provide a structural macroscopic view of it.

The presence of arches before and after the discharge allows relating the packing structure with the size of the density depression \cite{ref12}.

The higher the order in the initial structure, the greater the extent of the rarefaction area after discharge (Fig. \ref{figure7} (b) and Fig. \ref{figure8} (b)), i.e., a more open (less ordered) initial structure induces a less spreading of the lower density region.

Regarding the shape of the rarefaction region, the decrease in density is more pronounced upward and sideways to an angle of $60^{\circ}$ ($120^{\circ}$) (which defines the stagnation zone). This is evidenced by comparing the density profiles for the columns (Fig. \ref{figure8} (b)) and those for the circular sectors limited by straight lines at $60^{\circ}$ and $120^{\circ}$ (Fig. \ref{figure7} (b)) with respect to the case shown in Fig. \ref{figure5} (c).

After discharge in the case of Fig. \ref{figure7} (b), a sudden decrease in density is observed up to a height equivalent to the estimated Beverloo's diameter. Two regions can be distinguished: one corresponding to a radius of $2$ a.u., where the density is zero, and the other where the density increases rapidly, almost reaching the initial bulk value.

The results for the case $D=2.75$ are qualitatively the same as the one analyzed above, for that reason they are not presented here.

It is important to remember that here we consider a monosized distribution. Taking into account that in real applications the size distribution of particles is usually not monodisperse, it is a future challenge to see how our present results are modified when considering a given dispersion in the size distribution of particles.

\begin{acknowledgements}
This work was supported by CONICET (Argentina) through Grant PIP 353 and by the Secretary of Science and Technology of Universidad Nacional de San Luis, Grant P-3-1-0114.
\end{acknowledgements}

\end{document}